\documentclass[aps,prl,superscriptaddress,preprintnumbers,showpacs,twoside,twocolumn,floatfix]{revtex4}

\usepackage{bm}
\usepackage{graphicx}
\usepackage{amsmath}
\newcommand{\ket}[1]{\bigl\lvert#1\rangle }
\newcommand{\kla}[1]{\left( #1 \right)}

\begin{document}

\title{Localization of cold atoms in state-dependent optical lattices via a Rabi pulse}

\author{Birger Horstmann}
\affiliation{Max-Planck-Institut f\"ur Quantenoptik, Hans-Kopfermann-Stra\ss e 1, 85748 Garching, Germany}
\author{Stephan D\"urr}
\affiliation{Max-Planck-Institut f\"ur Quantenoptik, Hans-Kopfermann-Stra\ss e 1, 85748 Garching, Germany}
\author{Tommaso Roscilde}
\affiliation{Laboratoire de Physique - ENS Lyon, 46 All\'ee d'Italie, 69007 Lyon, France}

\begin{abstract}
We propose a novel realization of Anderson localization in non-equilibrium states of ultracold atoms in an optical lattice. A Rabi pulse transfers part of the population to a different internal state with infinite effective mass. These frozen atoms create a quantum superposition of different disorder potentials, localizing the mobile atoms. For weakly-interacting mobile atoms, Anderson localization is obtained. The localization length increases with increasing disorder and decreasing interaction strength, contrary to the expectation for equilibrium localization.
\end{abstract}

% 03.75.Lm - Tunneling, Josephson effect, Bose-Einstein condensates in periodic
%            potentials, solitons, vortices and topological excitations
% 03.75.Mn - Multicomponent condensates; spinor condensates
% 64.60.My - Metastable phases
% 72.15.Rn - Localization effects (Anderson or weak localization)

\maketitle

The control and manipulation of ultracold atoms in magnetic and optical traps has made enormous progresses towards the realization of fundamental condensed matter phenomena \cite{Zwerger07,Lewenstein07}. Among them, the phenomenon of quantum localization of matter waves induced by a random potential, observed in a variety of condensed-matter setups \cite{Andersonreview},  takes a new dimension in the context of cold atoms. In fact the capability of fine tuning the disorder potential as well as the interaction among atoms within the same sample leads to the possibility of investigating the Anderson localization transition with unprecedented control, and to monitor the complex effect of interparticle interactions \cite{Fallanietalreview}. Substantial progress has been made recently in this direction with optical disorder potentials, leading to the observation of Anderson localization for weakly-interacting bosons \cite{Billyetal08,Roatietal08, Deissleretal10} and to experiments
aiming at the realization of a strongly interacting Anderson insulator (Bose glass) with bosons in optical lattices \cite{Fallanietal07, Whiteetal09, Pasienskietal09}.

An alternative route to Anderson localization of cold atoms consists in creating a disorder potential by pinning a random distribution of an atomic species $b$. A second, mobile atomic species $a$ experiences a disorder potential due to interspecies collisions. This localizes species $a$ \cite{GavishC05,Paredesetal05, RoscildeC07, Horstmannetal07}. This approach has the unique advantage that the disorder potential can be prepared in a quantum superposition of different realizations, such that the time evolution of the mobile $a$-atoms samples all realizations of the potential at once, and measurements on the mobile particles deliver disorder-averaged observables \cite{Paredesetal05,Horstmannetal07}.

%One strategy to achieve this is the independent preparation of the $a$ and $b$ atoms, and subsequent switching of their mutual interaction \cite{Horstmannetal07}, which can be realized by controlling the spatial phase of different polarization components of the lattice \cite{Mandeletal03}. An alternative strategy is the diabatic loading of the two species in the same optical lattice, leading to metastable emulsion states which contain a disordered arrangement of $b$ atoms which localizes the $a$ atoms \cite{RoscildeC07}.

\begin{figure}[tb]
\includegraphics[width=55mm]{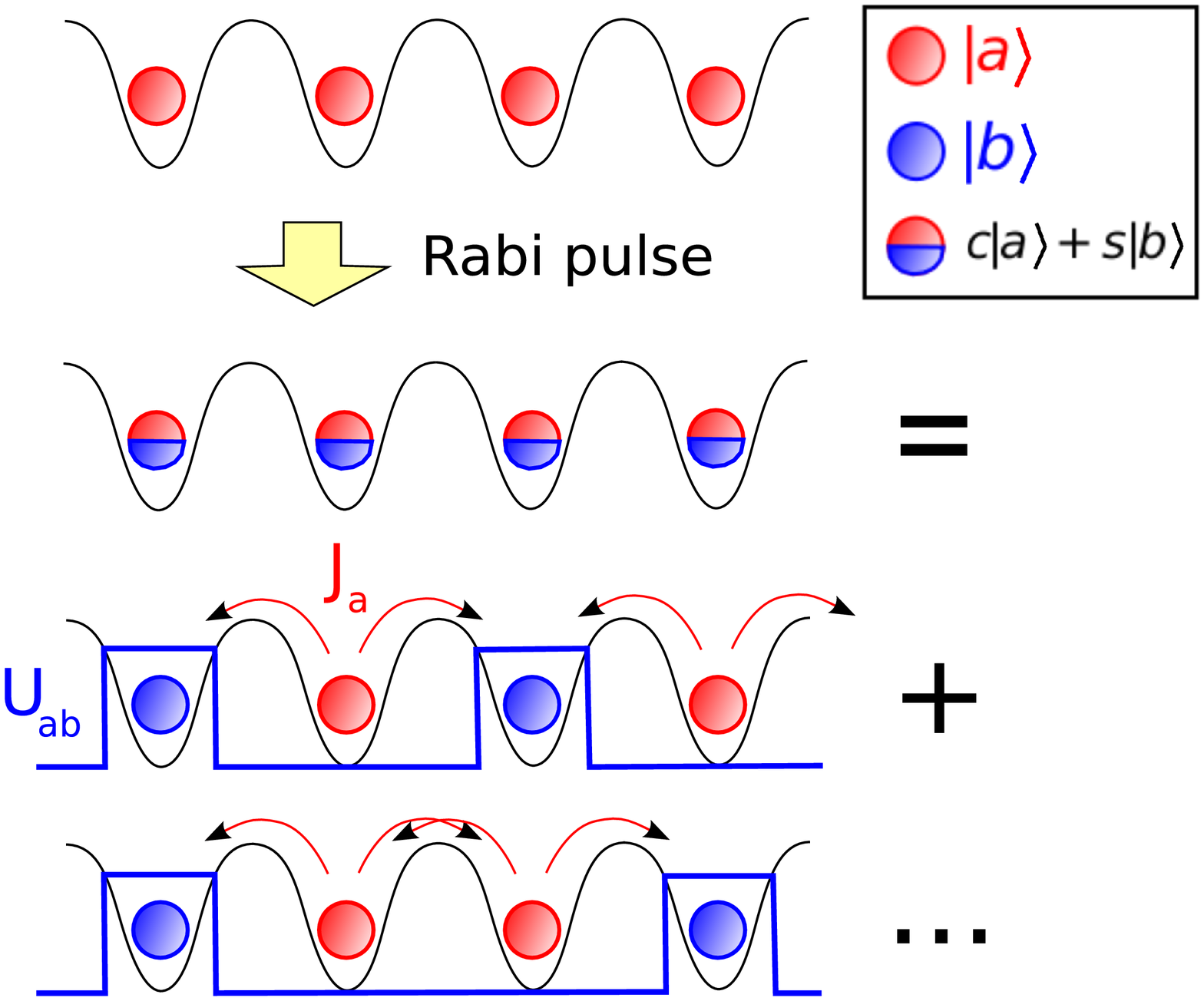}
%\vspace*{4.5cm}
\caption{Sketch of atom localization in state-dependent optical lattices. The $a$-atoms, initially prepared in the ground state of the lattice, are converted to a quantum superposition $\cos(\alpha/2)|a\rangle + \sin(\alpha/2)|b\rangle$ by a Rabi pulse, where the $b$-atoms are frozen by the lattice. The state of the system can be viewed as $a$-atoms (red) moving in a quantum superposition of random potentials created by the frozen $b$-atoms (blue).}
\label{f.sketch}
\end{figure}

In this paper, we theoretically explore a different scheme leading to localization in atomic mixtures, based on ``dynamical doping" of frozen $b$-atoms into a cloud of $a$-atoms. $a$ and $b$-atoms represent two internal states of the same atomic species. Like Refs.\ \cite{Paredesetal05, RoscildeC07, Horstmannetal07}, we use a state-dependent optical lattice to freeze the $b$-atoms and keep the $a$-atoms mobile. All $N$ atoms are initially prepared in state $\ket{a}$, and occupy the motional ground state. Subsequently, a short radio-frequency Rabi pulse is applied, creating an internal-state superposition $\cos (\alpha/2) |a\rangle + \sin (\alpha/2) |b\rangle$, where $\alpha$ is the pulse area. We show that the frozen $b$ component creates a quantum superposition of disorder potentials (see Fig.~\ref{f.sketch}). The wavefunction of weakly-interacting $a$-atoms in a two-dimensional lattice is studied numerically and evolves into a quasi-stationary profile with exponential tails. While we concentrate our discussion on bosons, our scheme should work equally well for fermions. The idea of dynamical doping can also generate new perspectives for other disordered systems.

The analysis of our scheme is interesting for several reasons: First, the fact that dynamical doping results in localization is not obvious, because the Rabi pulse transfers a fraction $\sin^2(\alpha/2)$ of the population at each site into state $\ket{b}$. Without measuring the $b$ distribution, no quantum projection noise is introduced and it is not \emph{a priori} evident why the $b$-atoms should generate disorder. Second, it is not obvious how
decoherence in the superposition between states $\ket{a}$ and $\ket{b}$ will affect localization. Third, we show that the resulting disorder potential
has negligible spatial correlations, unlike the most recent realizations in cold atom experiments \cite{Fallanietal07, Whiteetal09, Pasienskietal09, Billyetal08, Roatietal08, Deissleretal10}. Fourth, the experimental implementation of this scheme should be significantly simpler than the previously proposed schemes for disorder in atomic mixtures \cite{Paredesetal05, RoscildeC07, Horstmannetal07}.
These schemes involve either the diabatic preparation of metastable states, resulting in a disorder realization that is not deterministically controlled by the preparation scheme \cite{RoscildeC07};
or the individual preparation of each species on disjoint optical lattices, and the subsequent relative motion of the two
lattices to achieve coupling of the mobile particles to the disorder potential \cite{Paredesetal05, Horstmannetal07}.
Fifth, due to the quantum parallelism, single-shot experiments with large atom numbers directly provide disorder averages.

While our proposal is quite general, we wish to point out that an experimental implementation is feasible with existing technology in ultracold gases, in a way that all relevant parameters can be adjusted independently. One could, \emph{e.g.}, choose $^{87}$Rb atoms with hyperfine states $|a\rangle = |F=1, \linebreak[1] m_F=1\rangle$ and $|b\rangle=|F=2,m_F=2\rangle$. For circularly-polarized lattice light, there is a so-called magic wavelength near 787 nm where the $a$-atoms see no lattice potential \cite{Milleretal02}. The ratio of the tunneling amplitudes $J_a/J_b$ can be adjusted over a wide range by tuning the wavelength $\lambda$ near the magic wavelength. Hence, one can make $J_b$ very small, thus freezing the $b$-bosons. A Feshbach resonance \cite{Chin} for state $\ket{a}$ can be used to adjust the interaction strength $U_a$. Due to the state-selective character of Feshbach resonances, this leaves the interspecies interaction strength $U_{ab}$ unchanged. The intensity of the lattice light controls $U_{ab}/J_a$. Finally, this choice of states $\ket{a}$ and $\ket{b}$ is stable against spin exchange and it is only subject to very slow dipolar relaxation \cite{Pethick}.

\emph{The model} -- Our system consists of $a$ and $b$-bosons trapped in the lowest band of an optical lattice, described by
${\cal H}_{ab}  = {\cal H}_a + {\cal H}_b + U_{ab} \sum_i n_i^{a}  n_i^{b}$
with the Bose-Hubbard Hamiltonian
\begin{equation}
\label{H0}
{\cal H}_a  = -  J_a \sum_{\langle ij \rangle} \left(a_i^\dagger a_j+{\rm H.c.}\right)
+ \sum_i \left(\frac{U_a}{2} n^a_i\bigl(n^a_i-1\bigr) + V_i n^a_i\right)
\end{equation}
for the $a$-bosons, and the analogous ${\cal H}_b$ for the $b$-bosons. Here $a$, $a^{\dagger}$ are destruction/creation operators for the $a$-bosons, $n^{a} = a^{\dagger} a$, $V_i = \frac m2 \omega^2 \bigl({\bm r}_i-{\bm r}_0\bigr)^2$ is an overall atomic trap
of frequency $\omega/2\pi$, and $m$ is the atomic mass. We consider $U_{ab}>0$.

The system initially contains only the $a$-bosons in the ground state of the Hamiltonian ${\cal H}_a$, which can be expanded in the basis of lattice Fock states
\begin{equation}
|\Psi_a \rangle_0 = \sum_{\{n^a_i\}} C(\{n^a_i\}) \left( \prod_i \frac{1} {\sqrt{{n^a_i}!}} (a_i^{\dagger})^{n^a_i} \right) |0\rangle
\end{equation}
subject to the constraint $\sum_i n^a_i = N$.

Under the assumption that the Rabi pulse is fast compared to the Hamiltonian dynamics, the application of a pulse instantaneously rotates the bosonic operators at each site, $a_i^{\dagger} \to \cos (\alpha/2) a_i^{\dagger}  +  \sin (\alpha/2) b_i^{\dagger}$.
The resulting state can be expanded as a quantum superposition of many partitions of the $N$ atoms over species $a$ and \nolinebreak[4] $b$
\begin{equation}
|\Psi_{ab}\rangle_{t=0} = \sum_{\{n^{a}_i\},\{n^b_i\}} \tilde{C}\left(\{n^{a}_i\},\{n^b_i\}\right) |\{n^{a}_i\}\rangle \otimes |\{n^b_i\}\rangle
\label{e.expand}
\end{equation}
under the constraint $\sum_i \left( n^{a}_i + n^b_i \right) = N$. This state can be rewritten as
\begin{equation}
|\Psi_{ab}\rangle_{t=0} = \sum_{\{n^b_i\}} c(\{n^b_i\}) |\{n^b_i\}\rangle \otimes
|\Phi_a( \{n^b_i\}) \rangle_{t=0},
\label{e.after_rabi}
\end{equation}
under the constraint $\sum_i n^b_i \leq N$, where
\begin{equation}
|\Phi_a( \{n^b_i\})\rangle_{t=0} = c(\{n^b_i\})^{-1} \sum_{\{n^{a}_i\}} \tilde{C}\left(\{n^{a}_i\},\{n^b_i\}\right) |\{n^{a}_i\}\rangle
\label{e.psi.a}
\end{equation}
is the normalized state of the $a$-bosons subject to the constraint of having a Fock state $|\{n^b_i\}\rangle$ of the $b$-bosons. $c(\{n^b_i\})$ is a normalization constant and can be chosen to be real.

The subsequent time evolution of the system is governed by ${\cal H}_{ab}$. The evolved state at time $t$, $|\Psi_{ab}\rangle_t$, can again be expanded in the form of Eq.~\eqref{e.expand}. We now assume that tunneling of $b$-bosons is negligible, $J_b=0$, so that the $b$-bosons are frozen in the Fock state $|\{n^b_i\}\rangle$. Hence, the $a$-boson components in Eq.~\eqref{e.psi.a} evolve decoupled from each other as $|\Phi_a \{n^b_i\}) \rangle_t = \exp[-i{\cal H}_{ab}(\{n^b_i\})t/\hbar] \  \linebreak[1] |\Phi_a( \{n^b_i\}) \rangle_{t=0}$. At time $t$, observables $A_a$, referring to $a$-bosons only, take the expectation value \cite{Paredesetal05}%
\begin{equation}
\label{expectation value}
\langle A_a \rangle_t =  \sum_{\{n^b_i\}}\left|c \left(\{n^b_i\} \right)\right |^2
\null_t\langle \Phi_a \{n^b_i\}) |A_a| \Phi_a \{n^b_i\}) \rangle_t,
\end{equation}
which describes the average over parallel quantum evolutions of the $a$-bosons in all possible realizations of the static potential $V_i = U_{ab} n^b_i$ generated by the frozen $b$-bosons. These realizations follow the statistics dictated by the $|c(\{n^b_i\})|^2$ coefficients, which are completely controlled by the choice of the initial state of the $a$-bosons. In particular, if this state is homogeneously spread over a given region of space, the distribution of the resulting effective potential will be dominated by configurations in which a number $\sim N\sin^2(\alpha/2)$ of $b$-bosons is randomly scattered over this region, hence giving rise to an effective disorder potential.

With $J_b=0$, the Hamiltonian ${\cal H}_b$  obviously contributes only a phase factor to the time evolution of each $|\Phi_a \{n^b_i\}) \rangle_t$. This has no effect
on the expectation values in Eq.\ \eqref{expectation value}. Hence, ${\cal H}_b$ can be omitted in our numerical calculations below. Furthermore, if there is spin decoherence, \emph{i.e.}, if the relative phase of states $a$ and $b$ fluctuates due to experimental imperfections, this will cause fluctuations in the phases of all the $|\Phi_a \{n^b_i\}) \rangle_t$ in Eq.\ \eqref{e.after_rabi}. This also has no effect on the expectation values in Eq.\ \eqref{expectation value}. This insensitivity to spin decoherence is a strength of our scheme.

\emph{Disorder statistics} -- In the following, we consider weakly interacting ($U_a \lesssim J_a$) Bose gases. To analyze the statistical properties of the disorder, we describe the state before the Rabi pulse with a condensate wavefunction on a lattice \cite{Zwerger07}
\begin{equation}
|\Psi_a\rangle_{t=0}=\frac{1}{\sqrt{N!}}\kla{\sum_i \psi_i a_i^\dagger}^N\ket{0} =: |\{\psi_i\}\rangle^{\otimes N} ,
\label{e.condensate}
\end{equation}
which is a good approximation far from the superfluid-to-Mott-insulator transition (occurring for $U_a/J_a \sim 16$ with $n=1$). Here, $\psi_i = \Psi_i/\sqrt{N}$ is the effective single-particle state. After the Rabi pulse, the state can be written in the form of Eq.\ \eqref{e.after_rabi} with $\ket{\Phi_a\bigl\{n_i^b\bigr\}}_{t=0}=|\{\psi_i\}\rangle^{\otimes (N-N_b)}$ and
\begin{eqnarray}
\left|c\left( \{ n_i^b \} \right)\right|^2
&=& \kla{\sin^2 \alpha/2}^{N_b} \kla{\cos^2 \alpha/2}^{N-N_b}
\nonumber  \\
&& \times \frac{N!}{(N-N_b)!} \prod_i \frac{|\psi_i|^{2n_i^b}}{n_i^b !} .
\label{e.astate}
\end{eqnarray}
The $|c|^2$ coefficients describe the statistics of a random potential with an average number of frozen bosons per site $\langle n_i^b\rangle= \langle N_b\rangle \left|\psi_i\right|^2$ and an average total number of frozen bosons $\left< N_b \right>=N\sin^2(\alpha/2)$. The correlations in the disorder potential can be described by the density-density correlation function
\begin{equation}
g^{(2)}_{ij}=\frac{\langle b_i^\dagger b_j^\dagger b_j b_i \rangle}{\langle b_i^\dagger b_i \rangle\langle b_j^\dagger b_j\rangle}=1-\frac{1}{N}.
\end{equation}
Therefore, for large $N$ the disorder potential at different sites is uncorrelated. Thus in contrast to speckle potentials, our proposal does not suffer from problems with the autocorrelation in the disorder potential. The disorder distribution is, however, inhomogeneous because of the external trapping potential.

\emph{Numerical simulations} -- In our simulations, we take $N=50$ bosons, a two-dimensional $L\times L$ lattice with $L=60$ and $\frac m2 \omega^2 d^2= 0.004J_a$ which yields a density $\langle n\rangle\leq 1$ throughout the trap ($\langle n\rangle\approx 1$ for the $U_a = 0$ case). Here, $d=\lambda/2$ is the lattice spacing. We determine $|\psi_i|^2$ by numerically minimizing the Gross-Pitaevskii (GP) energy functional \cite{Pethick}
%\begin{equation}
%\hspace*{-.5cm}
%E = -J_a \sum_{\langle ij \rangle} \left(\Psi_i \Psi_j^{*} + {\rm c.c.} \right) +   \sum_i \left( \frac{U_a}{2} |\Psi_i|^2 + V_i\right)|\Psi_i|^2
%\end{equation}
subject to the condition that the total number of particles be fixed to $N$. Next, we generate disorder realizations $\{ n_i^b\}$ according to the $|c|^2$ statistics of Eq.\ \eqref{e.astate} via standard Monte Carlo Metropolis sampling \cite{Horstmannetal07}. Finally, we numerically solve the time-dependent GP equation with ${\cal H}_{ab}$, omitting ${\cal H}_b$ as discussed above.

\begin{figure}[tb]
\includegraphics[width=80mm]{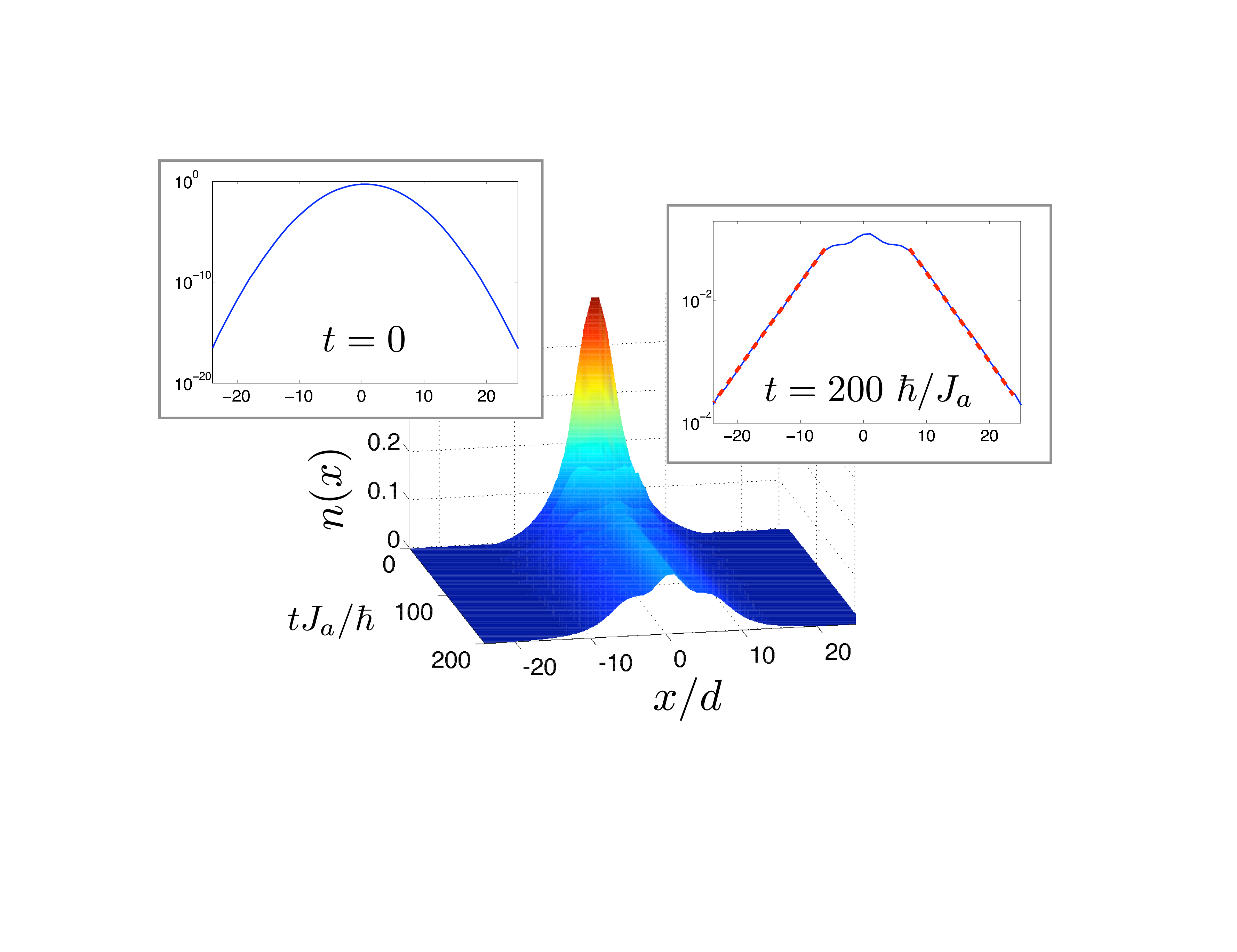}
\vspace*{-.5cm}
\caption{Time evolution of the density profile $n(x) = \langle n^a_{x/d} \rangle$ for the non-interacting gas, $U_a=0$, after a pulse of duration $\alpha = \pi/2$ and with interspecies interaction $U_{ab} = 7J_a$.}
\label{f.density}
\end{figure}

Fig.~\ref{f.density} shows the time evolution of the density profile of the $a$-bosons for $U_a = 0$, $U_{ab} = 7 J_a$, $\alpha = \pi/2$. After the Rabi pulse at $t=0$, the $a$-boson cloud rapidly expands from its initially Gaussian state, due to the repulsion from the dynamically doped $b$-bosons. The expanding cloud wings are then reflected by the trap. This generates breathing oscillations, which are quickly damped by the disorder. Remarkably the resulting, quasi-stationary density profile features a central peak with exponentially decaying tails, $\langle n_i \rangle \sim \exp(-|\bm r_i-\bm r_0|/\xi)$, clearly exhibiting Anderson localization. A similar behavior is observed for $U_a>0$ and for different values of $U_{ab}$ and $\alpha$. Yet, the dependence of the localization phenomenon on these parameters is nontrivial and differs significantly from what one expects at equilibrium.

\begin{figure}[tb]
%\vspace*{6.7cm}
\includegraphics[width=80mm]{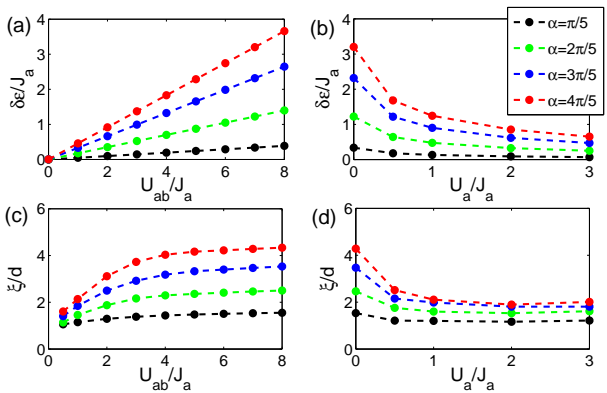}
%\mbox{
%\includegraphics[width=42mm]{localizationlengthvarU_less.eps}
%\includegraphics[width=40mm]{localizationlengthvarW_less.eps}}
%\null\hspace*{-.1cm}
%\mbox{\includegraphics[width=44mm]{EnergyDifferenceNormalizedvarU_less.eps}
%\includegraphics[width=42mm]{EnergyDifferenceNormalizedvarW_less.eps}}
\caption{(a),(b): transferred energy per $a$-boson (a) for a non-interacting gas of $a$-bosons ($U_a=0$), and (b) for interacting $a$-bosons with $U_{ab} = 7J_a$. (c),(d): corresponding localization length (in units of the lattice spacing $d$) of the evolved cloud.}
\label{f.xi}
\vspace*{-.4cm}
\end{figure}

The fast, diabatic Rabi pulse takes the system out of the initial ground state into an excited non-equilibrium state.
According to Eq.~\eqref{e.astate},
the remaining $a$-bosons occupy a superposition of different condensate states with
variable particle number $N-N_b$; these condensate states are non-equilibrium ones
because they are immersed in the disordered background of frozen $b$-bosons, and
because, in presence of finite interactions $U_a>0$, the condensate wavefunction
is no longer minimizing the interaction energy for a reduced particle number $N-N_b$.
To quantify the distance from equilibrium, we evaluate the energy transfer per $a$-boson due to the Rabi pulse
\begin{equation}
\delta \varepsilon  = \frac{ \langle {\cal H}_{a} + U_{ab} \sum_i n_i^a n_i^b \rangle_{\rm after}}{N\cos^2(\alpha/2)} -
\frac{\langle  {\cal H}_{a} \rangle_{\rm before}}{N} ,
\end{equation}
where $\langle...\rangle_{\rm before(after)}$ denotes the expectation value over the wavefunction before (after) the pulse. Fig.~\ref{f.xi}a/b clearly shows that the transferred energy $\delta \varepsilon$ increases for increasing disorder strength (namely for increasing $\alpha$ and $U_{ab}$), while it decreases for increasing interactions $U_a$.
This is partly due to the fact that, for a given initial distribution (controlled uniquely by the $U_a/J_a$ ratio and by the trapping potential), the gain in on-site interaction for the $a$-bosons is $\propto U_{ab} - U_a/2$, namely it increases linearly with $U_{ab}$. On the other hand, increasing $U_a$ reduces the probability that many $a$-bosons occupy the same site before the Rabi pulse; this latter fact reduces the probability that, immediately after the pulse, a site will be occupied by both $a$ and $b$-bosons, and this reduces $\delta \varepsilon$.

So on the one hand, the equilibrium physics suggests that stronger localization (smaller $\xi$) should occur for stronger disorder $U_{ab}$ and for weaker interactions $U_a$, due to reduced screening \cite{Andersonreview, Deissleretal10}. On the other hand, the non-equilibrium physics suggests the exact opposite, namely that stronger $U_{ab}$ and weaker $U_a$ both increase $\delta\epsilon$, so that the $a$-bosons expand more violently, making the wings
of the quasi-stationary density profile fall off more slowly, and therefore increasing the localization length $\xi$. The evolution of $\xi$ in Fig.~\ref{f.xi}c/d shows that the non-equilibrium aspects dominate in this competition. Note that, for larger $\xi$, the density profile exhibits exponential tails over a larger spatial range, which facilitates the experimental observation of the tails \cite{noteonsupport}.

\begin{figure}[h]
\includegraphics[width=50mm]{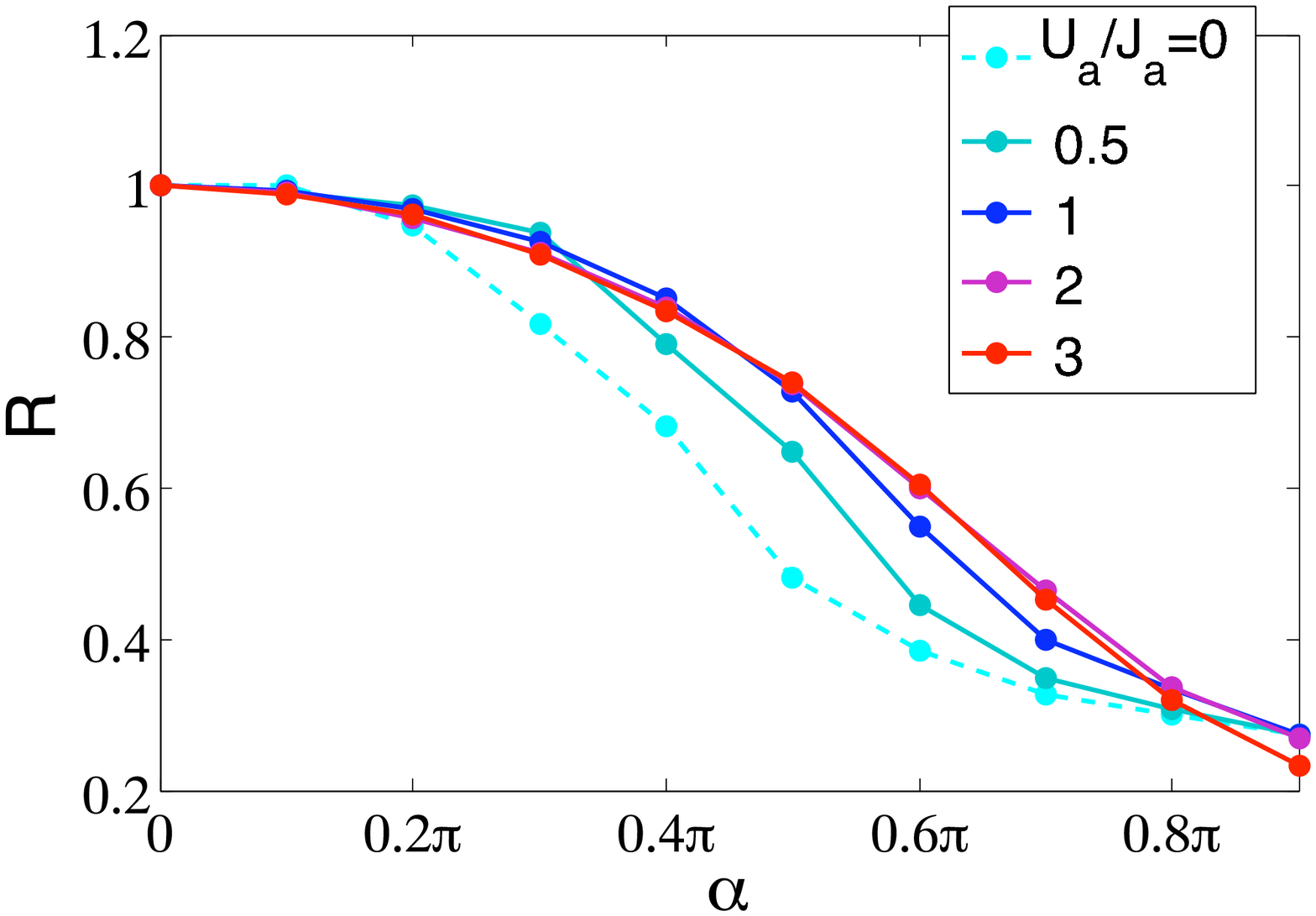}
%\vspace*{4.5cm}
\caption{Relative coherent fraction $R$ after time evolution (see text) for $U_{ab}=7J_a$.}
\label{f.nk0}
\end{figure}

While the real-space localization exhibits counterintuitive, non-equilibrium properties, the coherence properties of the expanding cloud exhibit instead a more intuitive behavior. We consider the momentum distribution of the $a$-bosons, $\langle n_k\rangle_t =\sum_{i,j}e^{-ik\kla{i-j}} \langle a_i^\dagger a_j\rangle_t/L^2$ right after the pulse ($t=0$) and after time evolution to a quasi-stationary state ($t=t_{\rm max}=200 \hbar/J_a$), and we monitor the loss of coherence during evolution via the relative coherent fraction $R = \langle n_{k=0}\rangle_{t_{\rm max}}/\langle n_{k=0} \rangle_{t=0}$. Fig.\ \ref{f.nk0} shows that the relative coherent fraction is suppressed for increased disorder strength (larger $\alpha$), showing that, although the cloud displays exponential tails with a longer $\xi$, the internal phase coherence is spoiled by disorder more effectively. This effect is reduced upon increasing the repulsion strength $U_a$, showing that disorder screening is indeed
observed in the cloud core, which is the part contributing more significantly to the overall coherence properties.

In conclusion, we have shown that novel, non-equilibrium disorder effects on the many-body state of bosons trapped in a spin-dependent optical lattice can be observed by dynamically doping frozen particles into the system via a Rabi pulse. The doped particles inherit the spatial structure and correlations from the initial state of the mobile particles, so that a rich disorder statistics can be generated by state preparation of the system before the pulse. Our predictions can be tested by state-selective time-of-flight measurements and \emph{in-situ} imaging of the bosonic cloud, already used to demonstrate Anderson localization of weakly interacting Bose gases in optically generated random potentials \cite{Billyetal08, Roatietal08, Deissleretal10}. Moreover an experimental implementation of dynamical doping is particularly desirable, as it could access the regime of strongly interacting mobile atoms, on the verge of the superfluid-to-Mott transition, which remains challenging
for all current theoretical approaches.

We thank D. Bauer, J. I. Cirac, and M. Lettner for discussions. This work was supported by the European Union via SCALA, by the German Excellence Initiative via the Nanosystems Initiative Munich, and by the DFG via SFB631.


\begin{thebibliography}{99}
\bibitem{noteonsupport} The spatial support of the disorder distribution is finite, being the same as that of the condensate wavefunction before the pulse. Part of the $a$-bosons can expand beyond this region, so that the extreme tails do not belong to the localized wavefunction, but to the wavefunction transmitted through the disorder.
\end{thebibliography}
\end{document}